# Open questions in massive star research

N. Langer

Universität Bonn and Max Planck Institut für Radioastronomie Bonn

**Abstract.** In discussing open question in the field of massive stars, I consider their evolution from birth to death. After touching upon massive star formation, which may be bi-modal and not lead to a zero-age main sequence at the highest masses, I consider the consequences of massive stars being close to their Eddington limit. Then, when discussing the effects of a binary companion, I highlight the importance of massive Algols and contact binaries for understanding the consequences of mass transfer, and the role of binaries in forming Wolf-Rayet stars. Finally, a discussion on pair instability supernovae and of superluminous supernovae is provided.

## 1. Introduction

For the task to get us started with this scientific conference on Massive Stars, which includes many different aspects of research, I decided to highlight what can be considered as important open questions in the field. I quickly realised that it is impossible to do this in a complete, or even in a comprehensive way. This contribution does therefore not attempt to review the state-of-the art or the recent progress in any of the topics discussed below. Specialists on the mentioned topics may be missing depth and detail. This paper is rather written for the non-specialists — which perhaps we all are in most aspects of massive star research.

The intention here is to identify and discuss unsolved problems which are significant not only for the listed topic but also for a wider scope. Clearly, there are many more of them than the ones brought forward here. While I tried to remain within the topical boundaries set by the conference program, it appears again impossible to cover all the contributing science fields, and it was also hard to fight my personal bias. I must therefore happen that readers feel important questions are missing here, and my hope is that they enjoy reading this paper nevertheless.

## 2. How do massive stars form?

The question of massive star formation is important because the answer sets the initial state for all of the further evolution. Any mistake in the assumptions on the initial state of massive stars may propagate. In stellar evolution calculations, mostly one of two different approaches are followed. In one of them, chemically homogeneous stellar models are initially put to hydrostatic and thermal equilibrium, on the so called zero age main sequence (e.g., Brott et al. 2011, Ekström et al. 2012). Alternatively, the models start with a fully convective structure in hydrostatic equilibrium on their Hayashi line, and further-on contract to eventually ignite hydrogen burning (e.g., Choi et al. 2016). Both approaches lead to nearly identical internal structures. However, they are pursued because they are simple, but they ignore at least two longstanding problems.

One problem is that young massive stars may still be accreting at the time when hydrogen burning ignites (cf., fig. 1 in Yorke, 1986). This issue may be related to the question why we see essentially no young massive stars (see Sect. 2.3). The second problem is that the vast majority of massive stars are born in binary or higher order multiple systems. When stars are born near the Hayashi line, they are big and do not allow for short orbital periods. We discuss this issue further in Sect. 2.1. For an in-depth review on massive star formation see Zinnecker & Yorke (2007).

### 2.1. *Which fraction of binaries merge as pre-main sequence stars?*

One may wonder why any pre-main sequence binaries should merge, because while after core hydrogen ignition the stars become bigger in size, the radii of pre-main sequence stars generally decrease (Siess et al. 2000). However, star formation occurs in a dense environment, and residual accretion as well as tides induced by circum-binary material may lead to shrinking orbits in pre-main sequence binaries (Korntreff et al. 2012). Based on a simple toy model which includes these processes, Tokovinin & Moe (2020) predict 33% of all early B type binaries to merge during their pre-main sequence evolution. Their model explains the observed B star binary fraction as well as the prevalence of short orbital periods in B type pre-interaction binaries (Villasenor et al. 2021).

A high fraction of pre-main sequence binary mergers implies a bi-modality in massive stars. One type would form via disk accretion, the other would gain a major fraction of their mass through a binary merger. The question is whether we can recognise this dichotomy in the long lived core-hydrogen burning state. One stellar quantity which may contain a memory of the formation process is the stellar spin. While merger products could have a large spin due to the surplus of orbital angular momentum (de Mink et al. 2013), Schneider et al. (2019, 2020), via detailed numerical simulations, find that outflows in the puffed-up post-merger state drain the merger product of angular momentum. Merger products could therefore be exceptionally slow rotators.

Indeed, in field stars, a bi-modal rotational velocity distribution has been found in early B (Dufton et al. 2013) and A type (Zorec & Royer 2012) stars. The same is true in rich, young open star clusters, where this phenomenon gives rise to a split main sequence band in high precision colour-magnitude diagrams (Milone et al., 2013, 2018; Marino et al. 2018). Star counts identify 10...30% of the main sequence stars as slow rotators. Here, the idea of different formation mechanisms of stars in the two main sequence components is supported by the fact that their mass functions are distinctly different (Wang et al. 2022).

### 2.2. *What causes large scale magnetic fields in stars?*

Between 5% and 10% of the early type main sequence stars has been found to host large scale magnetic fields (Grunhut et al. 2017, Schöller et al. 2017). While several formation pathways for these magnetic fields have been suggested (cf., Langer 2014), the idea that large scale, stable fields can be created in stellar mergers has recently gained momentum (Ferrario et al. 2009, Tutukov & Fedorova 2010, Schneider et. al. 2016, 2019, Pelisoli et al. 2022), and is supported by the low binary fraction of magnetic stars (Kochukhov et al. 2018).

However, further work is needed to verify this idea. The number of slow rotators amongst early B type stars, for example, seems clearly larger than the number of magnetic early B stars. Sticking to the merger idea, this could imply that not each merger generates a long-term stable B-field, perhaps because stability requires the intertwined toroidal-poloidal large scale topology found by Braithwaite & Spruit (2004), which may or may not form as remnant of the decay of an unstable turbulent or toroidal field generated during the merger. Alternatively, Fossati et al. (2016) suggested that large scale magnetic fields in massive stars may decay on a timescale comparable to their main sequence lifetime. Both scenarios have different consequences for the late time and post-core collapse evolution of massive stars, and magnetic stellar evolution models to explore them are highly demanded (Fuller et al., 2019, Keszthelyi et al., 2019, Takahashi & Langer 2021).

### 2.3. *Are young very massive stars embedded?*

During the last decade or so, large spectroscopic surveys of massive stars have shown a dearth of young stars above $\sim 20\,M_\odot$. This has been found in the Galaxy (Castro et al., 2014, Holgado et al., 2020), the LMC (Schneider et al. 2018), and the SMC (Schootemeijer et al., 2021) to comparable extent. Perhaps, the missing stars are still embedded in their birth clouds or even accreting (cf., Sect. 2.1). However, to solve the problem, they need to remain hidden at optical wavelengths for roughly half their total lifetime, but a corresponding population of IR bright stars seems also not observed.

Ongoing accretion during a fraction of their core hydrogen burning evolution would severely alter the internal composition profiles of massive main sequence stars. They would encounter the problem of rejuvenation, as the mass gainers in mass transferring close binaries (Braun & Langer, 1995; cf., Sect. 4.2), introducing a large uncertainty on their post-main sequence evolution.

## 3. Which effects has the Eddington limit?

Massive main sequence stars are known to follow a steeper than linear mass-luminosity relation, which means that above a certain mass limit they must approach or even exceed the Eddington limit. Usually, the Eddington limit is considered at the stellar surface, where it may have strong consequences for the stellar wind (Owocki et al. 2017). When evaluating the Eddington limit in the stellar envelope, Langer et al. (2012) showed that solar metallicity main sequence stars above $\sim 30\,M_\odot$ are expected to exceed it. However, instead of posing an upper mass limit, perhaps stars above the Eddington limit are inflated and variable.

### 3.1. *Do stellar envelopes inflate near the Eddington limit?*

A super-Eddington radiative luminosity in a stellar envelope does not imply the loss of hydrostatic equilibrium. Instead, in corresponding stellar structure models, a density inversion produces an inwards directed force compensating for the excess acceleration produced by the radiation. Ishii et al. (1999) investigated zero-age main sequence and helium main sequence models and found this effect to lead to inflated envelopes.

Petrovic et al. (2006) found inflated envelopes in models of massive Wolf-Rayet stars, and showed that the radius increase depends on the stellar wind mass loss rate. Gräfener et al. (2012), and Sanyal et al. (2015, 2017) showed the extensive degree of envelope inflation in models of evolved massive stars, where a radius increase of up to a factor of 40 has been found. Notably, the effects of density inversion and envelope inflation have also been retrieved in 3D-radiation-hydro models of Jiang et al. (2015), and are thus not an artefact of 1D models. We also need 2D-models of rotating stars near the Eddington limit in order to understand how the centrifugal force affects the envelope inflation.

### 3.2. *Are Luminous Blue Variable stars near their Eddington limit?*

The empirical verification of envelope inflation is difficult. The most direct way may relate to stability analyses of inflated models. Many inflated stellar models turn out to be pulsationally unstable (Grassitelli et al. 2016). However, the corresponding brightness fluctuations may be small and difficult to observe. On the other hand, Sanyal et al. (2015) found that models of evolved massive stars with a radius increase due to inflation of more than a factor of three where all found beyond the hot edge of the empirical LBV instability strip identified by Smith et al. (2004).

Indeed, Grassitelli et al. (2021) find strong radius and brightness variations resembling those of S Doradus type LBVs in time-dependant stellar evolution models which exhibit strong inflation. The variations occurred on a decade-long timescale, which is much larger than the dynamical timescale of those models, and relates to the thermal timescale of the inflated envelopes of these models. Grassitelli et al. showed that for plausible stellar wind mass loss recipes, these models can never reach a thermal equilibrium configuration, but keep undergoing cyclic large-amplitude variations. These models support the long-standing idea of a connection between LBVs and the Eddington limit (e.g., Lamers & Fitzpatrick, 1988).

Whether the great LBV eruptions à la $\eta$ Carina also relate to the Eddington limit is not clear yet. In a recent model, Hirai et al (2021) explain $\eta$ Car's great eruption as the result of a stellar merger in a very massive triple system. More 3D models of binary mergers (Schneider et al. 2019) are needed to systematically study their mass outflows, and to investigate their features when they occur near the Eddington limit.

## 4. Which effects has a binary companion?

The majority of massive stars is born with a binary companion in such a close orbit that Roche lobe overflow and mass transfer is unavoidable (Sana et al. 2012, Moe & Di Stefano, 2017). As mass is the single most important parameter of stellar evolution, mass transfer affects the evolution of both components drastically. Whereas the evolution of the mass donor can be predicted with large certainty (Sect. 4.3), the evolution of the mass gainer depends strongly on several uncertain physical processes, in particular on the mass transfer efficiency (Sect. 4.1) and on rejuvenation (Sect. 4.2).

### 4.1. *How efficient is mass transfer?*

When mass is transferred from the donor to the mass gainer, either by direct impact of the accretion stream (Dessart et al., 2003) or via the formation of an accretion disk (Lin & Pringle, 1976), either not all of the transferred matter may end up on the mass gainer or the mass gainer may immediately eject some of it. The mass transfer efficiency is the ratio between the amount of mass which remains on the mass gainer to the amount of mass transferred by the mass donor. While it is naturally a time dependant quantity (Langer 2012), often it is used as the time average over a mass transfer episode. Conservative evolution implies a mass transfer efficiency of one, and for non-conservative evolution, one needs to specify the angular momentum carried by any material which leaves the binary system in addition to the mass transfer efficiency.

As theoretical models for the mass transfer efficiency are sparse (e.g., Oka et al. 2002), in most evolutionary models of massive binaries a constant value for the mass transfer efficiency is adopted, which is then independent of time and of the binary parameters. This is problematic because we know that some binary systems evolved through mass transfer rather conservatively while other did not. E.g., some massive Algol binaries with accretor-to-donor mass ratios above three require near-conservative mass transfer (e.g., Sen et al. 2022), while we know WR+O star binaries which require highly non-conservative evolution (Petrovic et al., 2005). In recent large grids of detailed binary evolution models, we explored a mass transfer scheme in which the accretor stops to take in mass once it is spun-up to critical rotation (Langer 2012). This leads to very inefficient mass transfer for initially wide binaries in which the mass gainer is spun-up quickly (Packet 1981), but to rather conservative evolution in short period binaries where tides can prevent or slow down the spin-up. Detailed comparisons with the observed populations of massive Algols, Be stars, Be/X-ray binaries and Wolf-Rayet binaries imply that our mass transfer scheme is dissatisfactory for binaries with initial primary masses near $10\,M_\odot$, which produce most of the Be/X-ray binaries (see also Vinciguerra et al. 2020). On the other hand, it leads to results which agree well with observations for initial primary masses above $\sim 20\,M_\odot$, which produce most of the WR stars (Pauli et al., 2022) and black holes (Langer et al., 2020). A refined mass transfer scheme which accounts for the strong mass dependence of the mass transfer efficiency appears therefore warranted.

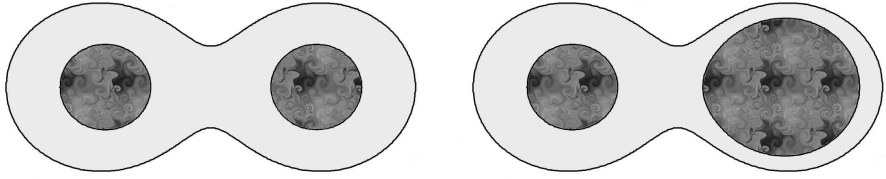

**Figure 1.** Schematic view of contact binaries in a thermal equilibrium state where the masses of both component stars, and thus their Roche volumes, are equal. In the left figure, complete rejuvenation is assumed, implying the same convective core masses and radii. The orbital period and the mass ratio (q=1) will not change further-on. In the right figure, it is assumed that rejuvenation did not occur, and therefore the core masses and radii are unequal, implying different timescales of envelope expansion for both components.

### 4.2. *Do mass gainers rejuvenate?*

As mentioned above, there is strong observational evidence for a high mass transfer efficiency in many binary systems. In other words, some mass gainers — for which we consider here only core hydrogen burning stars which is the most common case — increase their mass substantially. As more massive stars have more massive convective cores, accretion generally leads to an increase of the core mass. This implies an increase of the hydrogen mass fraction in the core, also called rejuvenation. To what extent mass gainers are rejuvenated is one of them most important and most longstanding unsolved questions in binary evolution. Often, it is assumed that the chemical composition profile of the mass gainer will adjust to that of a single star of the same mass and core hydrogen abundance (Hellings 1983). The problem is that any mixing of hydrogen into the core has to overcome the strongly stabilising mean molecular weight gradient. This process is likely controlled by semiconvection (Braun & Langer 1995), a thermal timescale mixing process of uncertain efficiency (e.g., Schootemeijer et al 2019). It turns out that contact binaries have the potential to elucidate the issue.

Menon et al. (2021) have computed a large grid of short period binary evolution models with MESA, which covers the initial parameter space of binaries which encounter a nuclear timescale contact phase, during which both stars overfill their Roche volumes (Fig.1). These models assumed conservative mass transfer, and efficient semiconvection and thus complete rejuvenation. When comparing to observations of massive contact binaries, Menon et al. found that many more contact binaries with a mass ration close to one were predicted than observed. For complete rejuvenation, when contact binaries encounter a mass ratio of one, there will be no further mass transfer, as both stars have the same internal structure, and thus expand at the same rate. Also the orbital period $P$ will not change any more, leading to very large predicted values of $P/\dot{P}$, which is again not observed (Abdul-Masih et al. 2022). While the models of Menon et al. are the first of their kind, the contact scheme in MESA does not account for a possible energy transfer inside the common envelope. I.e., both components of the MESA models can have different effective temperatures, whereas the observed temperatures of the components of contact binaries are similar (Abdul-Masih et al. 2021), as theoretically expected (Lucy 1968, Tassoul 2000). In any case, while this implies that both components of equal mass contact binaries are expected to have nearly identical luminosities and radii, they would only maintain a mass ratio of one if their internal structures were also identical, i.e., if complete rejuvenation would apply. On the other hand, a single $P/\dot{P}$-measurement well above the nuclear timescale would imply that complete rejuvenation works in nature, unless the observed binary was born with a mass ratio of one (Fig. 1).

In contact binaries, the von Zeipel theorem and Roche geometry also demand the luminosity ratio to equal the mass ratio of both components. This may be used as a consistency check in contact binary systems. In Fig. 2, we show the luminosity ratios in binaries which were observationally classified as contact binaries (large dots). While some are indeed close to the $L \sim M$-relation, others are very close to the $L \sim M^3$-relation, indicating they might be pre-interaction systems. Some systems are also found in between both relations, or even at luminosity ratios below one. This figure suggests that either the simple theoretical expectation is wrong, or that sometimes pre-interaction binaries with nearly Roche-lobe filling components or semi-detached Algol binaries are mistakenly classified as contact binaries. The accurate determination of the system parameters of a larger sample of such binaries will allow to settle this important question in the near future.

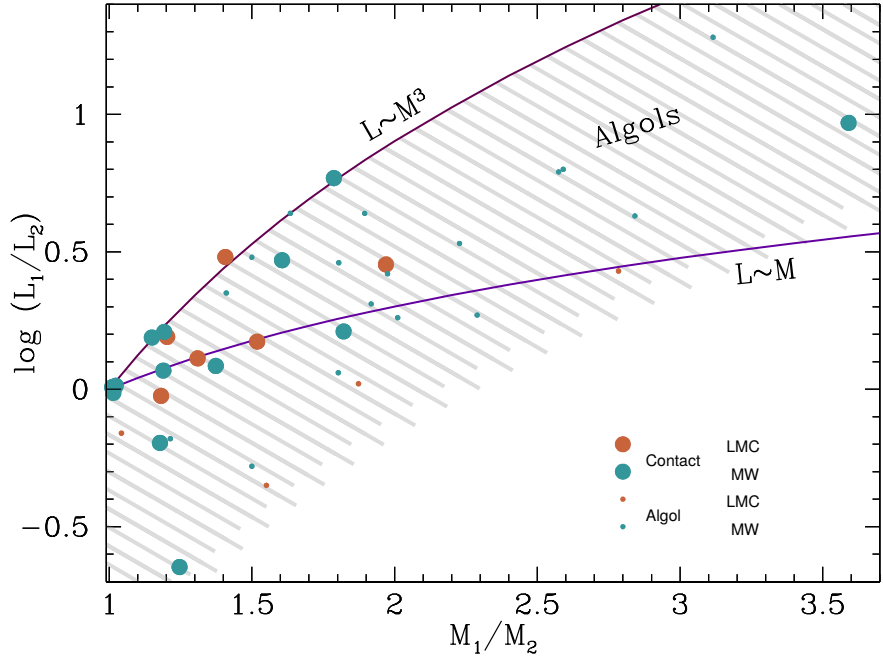

**Figure 2.** Mass and logarithmic luminosity ratios of the components of observed short period binaries. Subscript 1 refers to the more massive components, which are in the mass range $\sim 8\,M_\odot \ldots 40\,M_\odot$. The plot shows LMC (red) and MW (blue) systems, observationally classified as contact (large symbols) or semi-detached binaries (small symbols), as listed in tables 1 and 2 of Menon et al. (2021), and in tables 1 and 2 of Sen et al. (2022), respectively. The upper solid line indicates the locus of binaries in which the stars follow the single star mass luminosity relation, here assumed as $L \sim M^3$, es expected for pre-interaction binaries. The lower solid line indicates the relation $L \sim M$, which has been proposed to hold for contact binaries. The hatched area marks the mass-luminosity ratio range for Algol binaries (semi-detached) in the considered primary mass range as predicted by Sen et al. (2022; see their figure 13).

### 4.3. *Which fraction of the Wolf-Rayet stars were stripped by a companion?*

The question raised above is discussed controversially since long time (cf., T. Shenar, this volume). Single star evolution calculations have been explaining the properties of the observed WR populations with some success, in particular when strong rotationally induced mixing as well as strong stellar wind mass loss have been adopted (Maeder 1983, Langer et al. 1994, Meynet & Maeder 2005). However, it has been realised during the last decade that the majority of massive stars is born in close binary systems. As such, it appears unavoidable that binary stripping produces a significant fraction of the observed Wolf-Rayet stars.

In fact, in this respect, the predictions from binary evolution models are more robust than those from single star models. The mass donors are found to simply loose most of their H-rich envelope due to mass transfer, while the mass loss of single stars, in particular in the LBV and RSG stages, is still very uncertain. In a recent large binary model grid for LMC metallicity produced with MESA, Pauli et al. (2022) found the evolution of the mass donors to agree very well with comparable models produced with BPASS (Eldridge et al 2017). The models of Pauli et al. reproduce the number and luminosity distributions of the different types of Wolf-Rayet stars — which presumably from a complete population (Breysacher et al. 1999, Hainich et al. 2014) — to significant detail, without the need to include Wolf-Rayet stars produced by single stars.

A fraction of the observed WR stars may also originate from the initially less massive star in binaries. Once the initially more massive star has evolved via its Wolf-Rayet phase into a compact object, it may strip off the envelope of its companion either in a stable mass transfer phase or in a common envelope evolution. The fraction of Wolf-Rayet stars produced in this way is uncertain, mostly due to uncertainties in the formation kicks of compact objects, and in the common envelope physics. However, they may have black hole companions, and are good candidates to produce merging compact objects (van den Heuvel et al., 2017).

## 5. What is the origin of peculiar supernovae?

A large fraction of peculiar supernovae, and supernova subtypes, is likely originating from close binary evolution (e.g., Yoon et al. 2010, Zarpatas et al. 2019, Schneider et al. 2021). Sorting out the connections between observed supernovae and their progenitor evolution will keep refining our understanding of binary evolution. Here, I want to point out two questions relating to non-standard explosion mechanisms.

### 5.1. *Where are the pair-instability supernovae?*

In contrast to iron core collapse supernovae, where only about 1% of the available energy is transformed into the kinetic energy of the supernova, the predictions of pair-instability supernova (PISN) models have been extremely robust since the discovery of the instability (Rakavy & Shaviv 1967, Fraley 1968). There is no doubt that helium cores with masses above $\sim 60\,M_\odot$ will collapse before oxygen ignition, and produce a supernova due to explosive oxygen burning, as long as their mass is smaller than $\sim 135\,M_\odot$ (Heger & Woosley 2002, Heger et al. 2003), However, as yet, no observed supernova has been doubtlessly identified as PISN.

Obviously, the high required mass would make them rare events. In addition, at solar metallicity, stellar wind mass loss would likely reduce the final helium core mass well below 60 $M_\odot$, whatever the initial mass might have been (Yusof et al., 2022). On the other hand, stellar evolution models including the then most recent stellar mass loss prescriptions lead Langer et al. (2007) to conclude that PISNe could occur at metallicies of up to $Z_\odot/3$, which would result in one PCSN per 1000 core-collapse supernovae. Notably, based on more recent mass loss rate prescriptions, Higgins et al. (2021) find a PCSN metallicity threshold of only half of solar metallicity. There may be two ways to resolve this apparent contradiction.

Firstly, it is generally believed that PISNe are very bright and easy to spot (cf., Sect. 5.2). However, they may result from compact WR/He-stars, and up to He-core masses of 80 $M_\odot$ they produce essentially no radioactive $^{56}$Ni (Heger & Woosley 2002). E.g., Herzig et al. (1990) find a peak magnitude of only $M_V \simeq 14$ m in a PISN model from a 61 $M_\odot$ Wolf-Rayet star. Secondly, it is possible that a star undergoing a PISN is still covered by a massive and extended hydrogen envelope. This would drastically increase the peak brightness of the explosion. E.g., for such a case, with a helium core mass of 70 $M_\odot$, Kozyreva et al. (2014) find a peak magnitude of $M_V \simeq 19$ m. However, the resulting light curve strongly resembled that of a bright Type II-plateau supernova. Therefore, there might be PISNe not recognized as such amongst the observed Type II supernovae, which is are known to show a large diversity.

For the upper mass range of PISNe, i.e., for helium core masses above $\sim 100\,M_\odot$, the nickel masses produced in the explosion models exceed several solar masses. The corresponding explosions are extremely bright and fall into the regime of superluminous supernovae. Perhaps, those are restricted to the Early Universe (Whalen et al. 2013). Near the lower mass limit, the collapse induced by $e^{\pm}$-production becomes too weak to lead to explosive burning which can disrupt the star. Instead, in so called pulsational pair instability supernovae (which end up forming black holes) only parts of the envelope of the star may be ejected (Woosley 2017, Marchant et al. 2019). Observational evidence for these events is accumulating (Woosley et al., 2007, Mauerhan et al. 2013, Pastorello et al. 2013, Woosley & Smith 2022).

New supernova surveys may elucidate this question in the near future. They will also test the predictions of chemically homogeneous evolution in very massive binaries (de Mink et al., 2009), which leads to a viable channel to produce pairs of merging massive black holes (Marchant et al. 2016, Mandel & de Mink 2016). However, it is also producing pair instability supernovae, in the helium core mass range mentioned above. The cosmic population synthesis of these events by du Buisson et al. (2020) shows that even dim PISN should be detected with rates above one per year in surveys with magnitude limits of $\sim 22\,m$.

### 5.2. *What causes superluminous supernovae?*

During the last two decades, supernovae have been discovered which were more than 100 times more luminous that previously known ones, i.e., displaying $\sim 10^{51}$ erg in optical photons (cf., Quimby et al. 2011), called superluminous supernovae (SLSNe). Three different types of explanations for their high luminosity have been put forward. The first is a high amount of $^{56}$Ni, i.e., several solar masses, which seems possible only in the context of PISNe (Sect. 5.1). The second explanation assumes the supernova ejecta to ram into a dense wall of gas, i.e., solar masses of nearby circumstellar matter (as perhaps produced by pulsational pair instability supernovae). This breaks the supernova ejecta and converts their kinetic energy ($\sim 10^{51}$ erg for a typical supernova) into heat and photons. The third explanation assumes that the core collapse leading to the supernova produces an extremely rapidly rotating and extremely magnetised neutron star, a millisecond magnetar, which injects part of its rotational energy ($\sim 10^{52}$ erg) into the expanding supernova ejecta (Sukhbold & Woosley 2016).

For the most common type, the Type Ic SLSNe, the magnetar scenario appears to be favoured. Observationally, they appear to be related to the long-duration gamma-ray bursts, as both types of explosions have a preference to occur in low metallicity dwarf galaxies (Lunnan et al. 2014, Schulze et al. 2018), and possibly also to fast radio bursts (FRBs; Metzger et al. 2017). In an analysis of 38 Ic SLSNe, Nicholl et al. (2017) derive the required neutron star spins and magnetic field strengths within the magnetar model, as well as the ejecta masses and kinetic energies of the supernovae, which were found to correspond remarkably well to the parameters obtained from models of low-metallicity, chemically homogeneously evolving massive stars (Aguilera-Dena 2018, 2020), which have also been put forward as progenitors of long-duration gamma ray bursts (Yoon & Langer 2005, Woosley & Heger 2006, Yoon et al. 2006).

However, so far chemically homogeneously evolving stars have not yet been unambiguously identified, which leaves a big open question for this scenario. Furthermore, there are other types of SLSNe, most remarkably the hydrogen-rich SLSNe, for which the origin is not yet clarified (e.g., Kangas et al. 2022). With several large upcoming supernovae surveys in the UV, in the optical/IR, at radio wavelengths, and the resumption of the gravitational wave searches in the near future, it is likely that transient universe will keep to amaze us.

10    N. Langer